%% file: ms.tex
\documentclass[12pt]{iopart}
\pdfoutput=1

\usepackage{amssymb}
\usepackage{epsfig}
\usepackage{amscd}
\usepackage{graphicx,xspace}
\usepackage{pstricks}
\usepackage[left=3cm,top=3cm,right=3cm,bottom=3cm]{geometry}
\usepackage{color}

\usepackage{bm}

\usepackage{iopams}
\expandafter\let\csname equation*\endcsname\relax
\expandafter\let\csname endequation*\endcsname\relax
\usepackage{amsmath} 
\usepackage{amsthm}
\usepackage{amsfonts}
\usepackage{bbold}
\usepackage{bbm}
\usepackage{stackrel}

\begin{document}

\title[Strongly constrained stochastic processes: the multi-ends Brownian bridge]{Strongly constrained stochastic processes: the multi-ends Brownian bridge}

\author{Coline Larmier$^{1}$, Alain Mazzolo$^{1}$  and Andrea Zoia$^{1}$}

\address{$^{1}$Den-Service d'\'etudes des r\'eacteurs et de math\'ematiques appliqu\'ees (SERMA), CEA, Universit\'e Paris-Saclay, F-91191 Gif-sur-Yvette, France }
\ead{alain.mazzolo@cea.fr}
\vspace{10pt}

\date{Received:  / Accepted:  / Published }

\begin{abstract}
In a recent article, Krapivsky and Redner (J. Stat. Mech. 093208 (2018)) established that the distribution of the first hitting times for a diffusing particle subject to hitting an absorber is independent of the direction of the external flow field. In the present paper, we build upon this observation and investigate when the conditioning on the diffusion leads to a process that is totally independent of the flow field. For this purpose, we adopt the Langevin approach, or more formally the theory of conditioned stochastic differential equations. This technique allows us to derive a large variety of stochastic processes: in particular, we introduce a new kind of Brownian bridge ending at two different final points and calculate its fundamental probabilities. This method is also very well suited for generating statistically independent paths. Numerical simulations illustrate our findings.
\end{abstract}

%
\vspace{2pc}
%
%
\maketitle
%

\newcommand{\E}{\mathrm{E}}
\newcommand{\Var}{\mathrm{Var}} 
\newcommand{\Cov}{\mathrm{Cov}}

\newtheoremstyle{thmstyle}
	{9pt}
	{9pt}
	{\itshape}
	{}
	{\bfseries\scshape}
	{.}
	{\newline}
	{}
\theoremstyle{thmstyle}
\newtheorem{thm}{Theorem}



\section{Introduction}
\label{sec_intro}
Diffusion phenomena emerge in various fields of science, encompassing ecology~\cite{ref_book_Karlin}, reactor physics~\cite{ref_book_Pazsit} and finance~\cite{ref_book_Mikosch}. In many physical systems, the diffusive process evolves in a confined environment where boundaries play an important role~\cite{ref_book_Schuss}. Depending on the context, these boundaries can be of several kinds, reflecting, absorbing, semi-permeable, etc. Absorbing boundaries, corresponding to a process that is stopped (or killed) when reaching the frontier of a given domain, although largely studied in the literature~\cite{ref_book_Redner}, still reveal some surprising behavior. Indeed, very recently Krapivsky and Redner observed that the distribution of the first hitting times for a diffusing particle subject to hitting an absorber is independent of the direction of the external flow field~\cite{ref_Krapivsky}. The authors called this phenomenon the first-passage duality. Intrigued by this property, we would like to determine what kind of constraints (if any) could be strong enough so that the conditioned process does not anymore depend on the original drift. In this article, we establish sufficient conditions for such a phenomenon to arise. Our approach is based on conditioned diffusion~\cite{ref_book_Karlin}, also known as effective Langevin equations in the physics literature~\cite{ref_Majumdar_Orland,ref_Mazzolo_Jstat,ref_Mazzolo_JMP} or more formally conditioned stochastic differential equations in the mathematical literature~\cite{ref_Baudoin}. For instance, the case studied by Krapivsky and Redner where the diffusion is stopped when reaching a target corresponds, in our setting, to a diffusion conditioned by the first hitting time of a given level. Depending on the context, the conditioning could be of various kinds; however, the event (or events) on which the diffusion is constrained has usually a null probability. This is the case of the Brownian bridge, where the constraint of returning to zero (or any another value) at a fixed time has clearly a zero measure. Conditioning a subtle object like a diffusion on events of zero probability is not an harmless task~\cite{ref_Mazzolo_Jstat,ref_Mazzolo_JMP}, but this approach is a fruitful method since it sheds light on the important process studied by Krapivsky and Redner but also makes the simulations easy. The next step is to understand when the conditioning is so strong that it erases the original drift of the diffusion. The effective Langevin equation approach will allow us to answer this question, at least in the case where the diffusion is constrained to have a discrete and/or continuous law at a given time. Once this formalism is in place, we will recover many known constrained processes and explore new ones, including a Brownian bridge ending at two (or more) fixed values (with possibly different probabilities for each value).\\

The article is organized as follows: in Section~\ref{sec2} we derive the stochastic differential equation for a diffusion conditioned by its first passage-time. In Section~\ref{sec3}, we extend this approach to a more general context and establish the stochastic differential equation for a diffusion constrained to be distributed according to an arbitrary probability law at a fixed time. After introducing this formalism, we present a new kind of Brownian bridge ending at two different final points and we derive its fundamental probabilities in Section~\ref{sec4}, first by standard probabilistic tools and then by the martingale approach, a powerful technique that allows us to recover some of the previous results in a totally different (and elegant) way. Finally, Section~\ref{sec_Conclusion} presents some concluding remarks. Monte Carlo simulations illustrate our theoretical findings.

\section{Stochastic differential equation for an absorbed Brownian motion}
\label{sec2}

\subsection{General setting}
In this section, we derive the stochastic differential equation for a diffusion conditioned by its first passage-time.
Before starting, we briefly recall the results obtained by Krapivsky and Redner, focusing on the one-dimensional case for sake of simplicity. In this setting, let us consider the first hitting time of a level $a > 0$ by a diffusion, i.e., by a stochastic process that satisfies the stochastic differential equation (SDE):
\begin{equation}
\label{def_BM_with_drift_SDE}
   \left\{
       \begin{aligned}
	  dX_t & = \mu dt + \sigma dW_t   \\       
	  X_0  & = 0 \, ,
       \end{aligned}
   \right.
\end{equation}
\noindent where $W_t$ is a standard Brownian motion (Wiener process), $\mu$ the constant drift of the process and $\sigma >0$ its diffusion parameter. Let us denote $T_a$ this stopping time:
\begin{equation}
\label{def_stopping_time}
	T_a = \inf_{t \ge 0} \{t, X_t = a \} \, .
\end{equation}
\noindent For a positive constant drift $(\mu > 0)$, directed towards $a$, the process will surely hit this level. However, for a negative constant drift $(\mu < 0)$, directed away from $a$, the probability that the particle eventually hits the level $a$ is $e^{-2 \vert\mu\vert a /\sigma^2} < 1$, meaning that there is a strictly positive probability that the diffusion will never touch $a$~\cite{ref_book_Redner}. For the negative drift case, Krapivsky and Redner showed that the distribution of the first hitting time $T_a$ conditioned on the event that the particle reaches the level $a$ is the same as in the case of a positive drift. The distribution of the first hitting time is therefore independent of the sign of the drift $\mu$. The authors called this intriguing property the first-passage duality (in one dimension). In the next two paragraphs, we will study diffusions conditioned to hitting a fixed level $a$ from the stochastic dynamics point of view, an approach that is also very well suited for numerical simulations. We start by studying the driftless case (pure Brownian motion) and then we will consider the Brownian motion with constant drift.

\subsection{Driftless case}
Consider a Brownian motion with a constant diffusion parameter $\sigma > 0$, and a positive level $a > 0$. We want to condition the process with respect to its first hitting time $T_a$. The probability that a Brownian motion reaches a level $a$ for the first time at a given time is an event of probability zero. As mentioned in~\cite{ref_Mazzolo_Jstat,ref_Mazzolo_JMP}, conditioning with respect to a set of sample paths of probability zero requires special care. Despite the technical complexities generated by conditioning with such events of zero measure, the resulting diffusion is indeed well defined. Basically, there are two ways to achieve such conditioning. The first method consists in approximating the Brownian motion by a series of random walks, while the second technique consists in approximating the conditioning event. The latter is known as the Doob's {\it{h}}-transform~\cite{ref_book_Doob}. A simple presentation of this technique is provided in the book of Karlin and Taylor~\cite{ref_book_Karlin} (although without referring to Doob's name). This approach is also outlined from a physicist point of view in the recent article~\cite{ref_Majumdar_Orland}. The random walk approximation is well suited when the process is symmetrical, as the standard (driftless) Brownian motion. Since we will also condition on Brownian motion with drift, in this article we will always use Doob's technique.\\

\noindent The key ingredient of Doob's method is the following. Consider a diffusion process $\{X_t, 0 \le t \le T\}$ characterized by a drift $\mu(x)$ and a variance $\sigma^2(x)$. The process $X_t$ thus satisfies the stochastic differential equation, 

\begin{equation}
\label{diffusion_general}
  dX_t = \mu(x) dt+ \sigma(x) dW_t \, ,
\end{equation}
\noindent with the initial value $X_0 =0$. Now, let $\{X^*_t, 0 \le t \le T\}$ be the process conditioned on an event $G(T)$ between two times 0 and $T$: for instance, for a Brownian bridge the constraint $G(T)$ is the event $\{X_T=0\}$,  while for the case presently studied it is the event $\{X_t < a \mathrm{~for~} 0 \le t < T \mathrm{~and~} X_T=a\}$. Then, the drift $\mu^*(x,t)$ and the variance $\sigma^{*2}(x,t)$ of the constrained process are given by~\cite{ref_book_Karlin}:

\begin{equation}
   \left\{
       \begin{aligned}
       \sigma^*(x,t) & = \sigma(x) \, ,  \\
       \mu^*(x,t)    & = \mu(x) + \frac{ \sigma^2(x)}{\pi(x,t; G(T))}   \frac{\partial \pi(x,t; G(T))}{\partial x}  \, ,
       \end{aligned}
   \right.
\end{equation}

\noindent where $\pi(x,t; G(T))$ is the probability that, from the state value $x$ at time $t$, the sample path of $X_t$ satisfies the desired constraint $G(T)$ at time $T$. This result is obtained by showing that $\pi(x,t; G(T))$ satisfies an appropriate backward partial differential equation~\cite{ref_Majumdar_Orland,ref_Orland}. The previous equations show that: (i) the variance is not affected by the conditioning procedure, (ii) the drift of the conditioned process includes an extra term that forces the process to satisfy the constraint $G(T)$. Note that the new drift may be time-dependent and discontinuous~\cite{ref_book_Karlin} and requires the knowledge of the probability $\pi(x,t; G(T))$. Once this quantity is derived, Doob's technique can be successfully applied to various kinds of conditioned processes~\cite{ref_book_Karlin,ref_Majumdar_Orland,ref_Baudoin,ref_Orland,ref_Szavits,ref_Chetrite}. For a regular (driftless)  Brownian process, the two previous equations reduce to (for the ease of notation we drop the $G(T)$ everywhere)

\begin{equation}
\label{drift_constrained_process}
   \left\{
       \begin{aligned}
       \sigma^*(x,t) & = \sigma \, ,  \\
       \mu^*(x,t)    & = \frac{\sigma^2}{\pi(x,t)} \frac{\partial \pi(x,t)}{\partial x} \, .
       \end{aligned}
   \right.
\end{equation}
\noindent In the present case, the probability $\pi(x,t)$ can be obtained by closely following the procedure described in~\cite{ref_book_Karlin} for the Brownian bridge. For this purpose, we define $\pi_{\epsilon}(x,t)$ as the (strictly positive) probability:

\begin{equation}
\label{def_pi_epsilon}
	\pi_{\epsilon}(x,t) = \mathrm{Prob} \left[\{X_s < a \mathrm{~for~}  t \le s < T_a \} \mathrm{~and~} \{X_{T_a} \in [a-\epsilon;a]\}\vert \, W_t = x \right] \, ,
\end{equation}
\noindent which is the probability that a Brownian motion $X_t$ starting at $x$ at time $t$ will remain under $a$ during the time interval $[t,T_a)$ and reach the small interval $[a-\epsilon;a]$ at time $T_a$. Then, we will get $\pi(x,t)$ 
by allowing $\epsilon$ to go to zero.
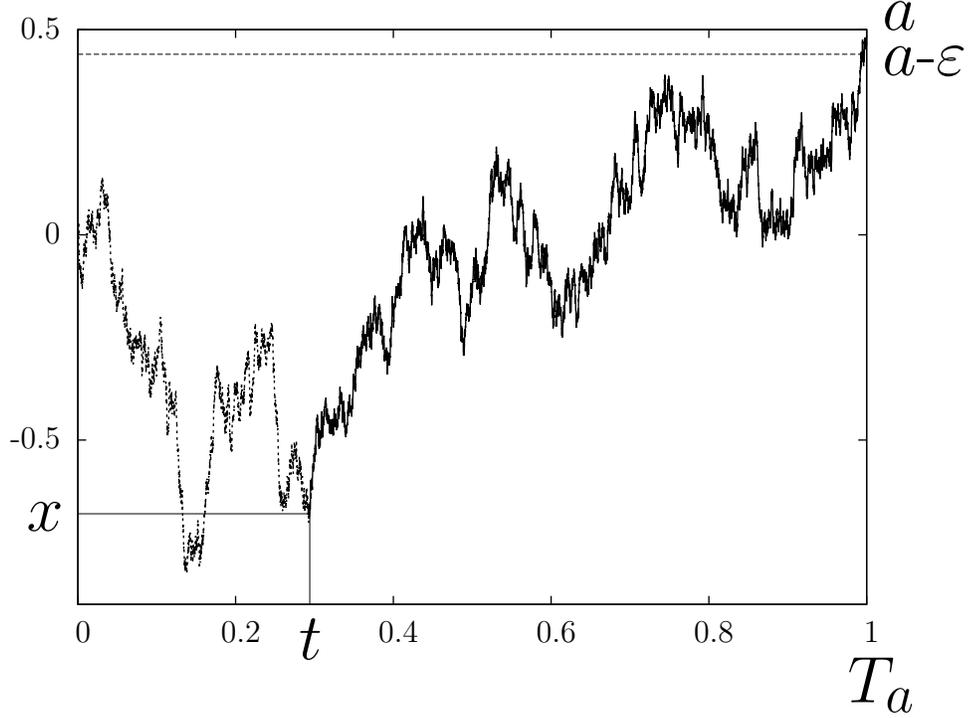
\begin{figure}
\begin{center}
\scalebox{1.0}{\input{brownianhitting}}\,\,\,\,
\end{center}
\caption{An example of realization for calculating the probability $\pi_{\epsilon}(x,t)$. The process starts at $x$ at time $t$, stays under the level $a=0.5$ until it lies between $[a-\epsilon;a]$ at time $T_a = 1$.}
\label{fig1}
\end{figure}

\begin{figure}[h]
\centering
\includegraphics[width=5in,height=4.in]{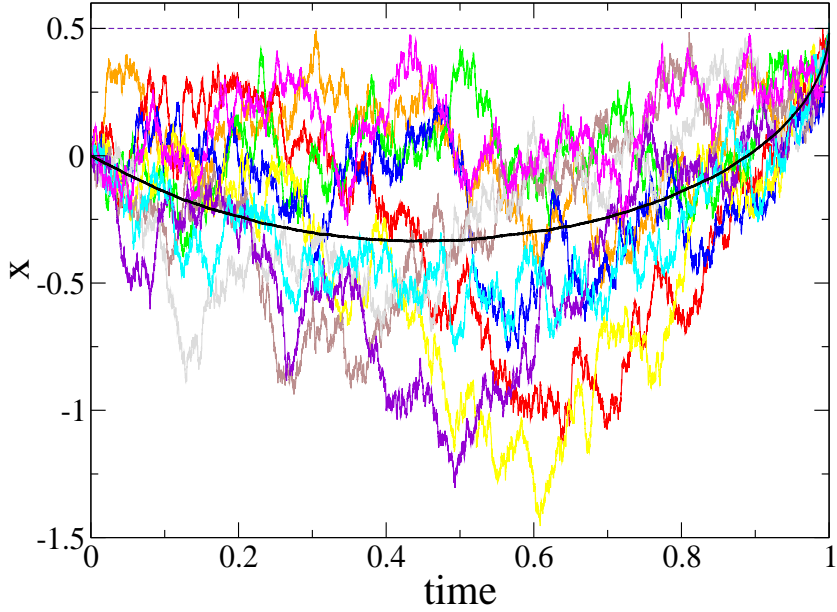}
\setlength{\abovecaptionskip}{15pt} 
\caption{A sample of 10 diffusions ending at $a=0.5$ at time $T_a = 1$ and conditioned to remain under the threshold $a=0.5$ for $t < T_a$. The time step used in the discretization is $dt=10^{-4}$. All trajectories generated with different noise histories are statistically independent. The thick black curve is the average profile of the stochastic process obtained by averaging over $10^4$ sample paths.}
\label{fig2}
\end{figure}

We are now ready to obtain $\pi_{\epsilon}(x,t)$ (and therefore $\pi(x,t)$) by direct calculation. Since the process is stopped when reaching the level $a$ for the first time, $a$ is an absorbing boundary. For such a process, starting in $x_0$ at time $0$, the concentration $c(x,t|x_0) = \E\left[X_t \in [x;x+dx] | X_0 = x_0 \right]$ is well known and can be obtained thanks to the method of images~\cite{ref_book_Redner}, namely,

\begin{equation}
	c(x,t|x_0) = \frac{1}{\sqrt{2 \pi \sigma^2 t}} \left( e^{- \textstyle \frac{(x- x_0)^2}{2 \sigma^2 t}} - e^{- \textstyle \frac{(x- 2a + x_0)^2}{2 \sigma^2 t} } \right)  \, .
\end{equation}
\noindent Since the particle may be absorbed, the integral of the concentration over the domain may be less than one. The probability distribution function $p(x,t|x_0)$ is then obtained by normalizing the concentration, namely, $p(x,t|x_0)=c(x,t|x_0)/\int c(x,t|x_0) dx$. Note that the normalization is not mandatory since only the logarithmic derivative of $\pi(x,t)$ is involved in Eq.\eqref{drift_constrained_process}. For sake of completeness.

\begin{equation}
	p(x,t|x_0) = \frac{1}{\sqrt{2 \pi \sigma^2 t}} \frac{1}{\mathrm{erf} \displaystyle{\left( \frac{a-x_0}{\sqrt{2 \pi \sigma^2 t}} \right)}} \left( e^{- \textstyle \frac{(x- x_0)^2}{2 \sigma^2 t}} - e^{- \textstyle \frac{(x- 2a + x_0)^2}{2 \sigma^2 t} } \right) ,
\end{equation}
\noindent where $\mathrm{erf}(x)$ is the error function. The probability $\pi_{\epsilon}(x,t)$ given by Eq.\eqref{def_pi_epsilon} is then obtained by integrating the previous equation over the interval $[a-\epsilon;a]$ ($X_t =x$ is now the starting point and the time interval is $T_a-t$, as shown in Fig.\ref{fig1}), namely,

\begin{equation}
	\pi_{\epsilon}(x,t) = \frac{1}{\sqrt{2 \pi \sigma^2 (T_a-t)}} \frac{1}{\mathrm{erf} \displaystyle{\left( \frac{a-x}{\sqrt{2 \pi \sigma^2 (T_a-t)}} \right)}} \displaystyle {\int_{a-\epsilon}^{a} }\left( e^{- \textstyle \frac{(y- x)^2}{2 \sigma^2 (T_a-t)}} - e^{- \textstyle \frac{(y- 2a + x)^2}{2 \sigma^2 (T_a-t)} } \right) \, dy \, .
\end{equation}
\noindent Then, we get
\begin{equation}
       \mu^*(x,t)     = \lim_{\epsilon \to 0} \frac{\sigma^2 }{\pi_{\epsilon}(x,t)}\frac{\partial \pi_{\epsilon}(x,t)}{\partial x} =  \frac{\sigma^2}{x-a} + \frac{a-x}{T_a-t} \, ,
\end{equation}
\noindent so that the constrained process $X^*_t$ satisfies the stochastic differential equation
\begin{equation}
\label{SDE_driftless}
	dX^*_t = \left(\frac{\sigma^2}{X^*_t-a} + \frac{a - X^*_t}{T_a-t} \right) dt + \sigma dW_t \, .
\end{equation}
\noindent In the mathematical literature, Eq.\eqref{SDE_driftless} (with the usual convention $\sigma=1$ and $T_a=1$) is obtained thanks to the (initial) enlargements of filtration technique~\cite{ref_Baudoin,ref_book_Mansuy}. From Eq.\eqref{SDE_driftless}, the corresponding Langevin equation follows immediately, i.e.,

\begin{equation}
\label{Langevin_driftless}
 \frac{dX^*_t}{dt} = \frac{\sigma^2}{X^*_t-a} + \frac{a - X^*_t}{T_a-t} + \sigma \eta_t,
\end{equation}
\noindent where $\eta_t$ is a Gaussian white noise process~\cite{ref_Majumdar_Orland}. Figure~\ref{fig2} shows a set of 10 realizations of the process conditioned to remain under the level $a = 0.5$ and having its first hitting time at $T_a=1$.
\noindent When the time $T_a$ becomes arbitrarily large ($T_a \to \infty$), the stochastic differential equation Eq.\eqref{SDE_driftless}  reduces to that of the taboo process, 
\begin{equation}
\label{SDE_taboo}
	dX^*_t = \frac{\sigma^2}{X^*_t-a} dt + \sigma dW_t \, ,
\end{equation}
\noindent namely a Brownian motion conditioned to remain forever below a certain threshold, which was originally introduced by Knight in one dimension~\cite{ref_Knight}. See also~\cite{ref_Pinsky} for a presentation of taboo processes in bounded domains or~\cite{ref_Mazzolo_Taboo} for a recent physicist-oriented survey. \\

\noindent When the level $a$ becomes large ($a \to \infty$), the first term in the r.h.s. of Eq.\eqref{SDE_driftless} is small compared to the second term, except when $X^*_t$ approaches $a$ close to the final time. In this case, the SDE Eq.\eqref{SDE_driftless} becomes

\begin{equation}
\label{SDE_Brownian_bridge}
	dX^*_t \underset{a \uparrow}{\sim}  \frac{a - X^*_t}{T_a-t} dt + \sigma dW_t \, , 
\end{equation}
\noindent which is the SDE of a Brownian bridge ending at $a$ at the final time $T_a$~\cite{ref_book_Karlin,ref_Majumdar_Orland}. This can be understood intuitively since, when $a$ is large, the process spends most of the time far from the boundary (recall that it starts at $x_0 = 0 \ll a$) and thus does not feel the frontier, except at the final time when the process is constrained to end at the level $a$. Apart from near-final times, the process therefore has a very low probability of being above $a$. This heuristic argument is confirmed by simulations.



\subsection{Brownian motion with constant drift}
In a similar way, we now condition a diffusion $X_t$ with constant drift $\mu$ and variance $\sigma$, that is  
\begin{equation}
\label{diffusion}
  dX_t = \mu dt+ \sigma dW_t \, .
\end{equation}
\noindent The parameters of the constrained process are given by 
\begin{equation}
\label{drift_BM_with_drift}
   \left\{
       \begin{aligned}
       \sigma^*(x,t) & = \sigma \, ,  \\
       \mu^*(x,t)    & = \mu + \frac{\sigma^2}{\pi(x,t)} \frac{\partial \pi(x,t)}{\partial x}  \, .
       \end{aligned}
   \right.
\end{equation}
\noindent where $\pi(x,t)$ is defined as in the previous section. Again, the concentration $c(x,t|x_0)$ for a process starting at $x_0$ at time $t_0 = 0$ can be obtained via the method of images~\cite{ref_book_Redner}, i.e.,
\begin{equation}
	c(x,t|x_0) \propto \frac{1}{\sqrt{2 \pi \sigma^2 t}} \left( e^{- \textstyle \frac{(x- x_0 -\mu t)^2}{2 \sigma^2 t}} - e^{- \textstyle \frac{2 \mu}{\sigma^2} (x_0 -a) }  e^{- \textstyle \frac{(x- 2a + x_0 - \mu t)^2}{2 \sigma^2 t} } \right) \, ,
\end{equation}
\noindent from which we get the probability

\begin{equation}
	\pi_{\epsilon}(x,t) \propto \displaystyle {\int_{a-\epsilon}^{a} } \left( e^{- \textstyle \frac{(y- x -\mu (T_a-t))^2}{2 \sigma^2 t}} - e^{- \textstyle \frac{2 \mu}{\sigma^2} (x -a) }  e^{- \textstyle \frac{(y- 2a + x - \mu (T_a-t))^2}{2 \sigma^2 t} } \right) \, dy .
\end{equation}
\noindent The limit follows straightforwardly,
\begin{equation}
       \lim_{\epsilon \to 0} \frac{\sigma^2}{\pi_{\epsilon}(x,t)} \frac{\partial \pi_{\epsilon}(x,t)}{\partial x} = \frac{\sigma^2}{x-a} + \frac{a -x -\mu (T_a-t)}{T-t} \, ,
\end{equation}
\noindent so that we have the drift
\begin{equation}
       \mu^*(x,t)     = \mu + \lim_{\epsilon \to 0} \frac{\sigma^2}{\pi_{\epsilon}(x,t)} \frac{\partial \pi_{\epsilon}(x,t)}{\partial x}=  \frac{\sigma^2}{x-a} + \frac{a-x}{T_a-t} \, .
\end{equation}

\noindent The stochastic differential equation satisfies by the conditioned process is thus
\begin{equation}
\label{SDE_with_drift}
	dX^*_t = \left(\frac{\sigma^2}{X^*_t-a} + \frac{a - X^*_t}{T_a-t} \right) dt + \sigma dW_t ,
\end{equation}
\noindent and does not depend of the original constant drift $\mu$. Equation\eqref{SDE_with_drift} is exactly the same as in the driftless case Eq.\eqref{SDE_driftless}. At this stage, since both constrained stochastic differential equations for the drifted and driftless cases are the same, it would be tempting to boldly conclude that the distribution of the first hitting time does not depend of the original drift $\mu$. This is obviously wrong, since for the case with drift the first hitting time is given by~\cite{ref_Krapivsky,ref_book_Redner} 

\begin{equation}
\label{distribution_Ta}
	P(T_a) = \frac{a}{\sqrt{2 \pi \sigma^2 T_a^3} }  e^{-\frac{(a-\vert\mu\vert T_a)^2}{2 \sigma^2 T_a}} \, .
\end{equation}
\noindent What is hidden in the stochastic differential equations Eqs.\eqref{SDE_driftless} and \eqref{SDE_with_drift}  is the fact that $T_a$ is not a fixed horizon time, but a random time. There is no contradiction between our approach and the first-passage duality. Indeed, the effective Langevin approach suggests the following: sample a random time $T_a$ according to Eq.\eqref{distribution_Ta}, then the dynamics of the constrained stochastic process is given by Eq.\eqref{SDE_with_drift} and thus does not depend of the original drift. The SDE approach is not an alternative proof of the first-passage duality property but gives instead complementary information. Besides, it raises the important question of knowing when a conditioning is strong enough so that the SDE of the conditioned process does not depend of the original diffusion drift at all. \\

\noindent To the best of our knowledge, apart from a few special cases~\cite{ref_Baudoin,ref_book_Mansuy,ref_Baudoin_finance}, there is no general theory concerning the conditioning of a Brownian motion (or a diffusion) by random times. To avoid the subtleties involved by such a procedure, from now on we will always condition the original diffusion to a fixed horizon time $T$.

\section{SDE for a Brownian motion with constant drift conditioned to have an arbitrary distribution at a fixed time}
\label{sec3}
In this section, we assume that $X_t$ is an unconstrained Brownian motion with constant drift $\mu$, and we wish to condition the process to an horizon time $T$. More precisely, we want the process to be distributed according to an arbitrary probability distribution function $f$ at time $T$. This probability function can be discrete, continuous or involving both a discrete and a continuous part (in other words, it can be any measurable function). At time $t < T$ the unconditioned process has a density 

\begin{equation}
		p(x,t) = \frac{1}{\sqrt{2 \pi t \sigma^2}} e^{ \textstyle - \frac{(x- \mu t)^2}{2 t \sigma^2}} \, ,
\end{equation}
\noindent for $-\infty < x < +\infty$. We wish to establish a correspondence between this density and the desired probability function $f(x)$ at the final time $T$. Again, this is achieved through the probability $\pi(x,t)$ that, from the state value $x$ at time $t$, the sample path of $X_t$ satisfies the desired constraint (here a given probability function $f$) at time $T$. The transition probability of the Brownian motion with constant drift from $(x,s)$ to $(y,t)$ is
\begin{equation}
		p(x,s;y,t) = \frac{1}{\sqrt{2 \pi (t-s) \sigma^2}} e^{ \textstyle - \frac{(x-y- \mu (t-s))^2}{2 (t-s) \sigma^2}} \, ,
\end{equation}
\noindent with $t>s$. By the Bayes rule, the transition probability $p^*(x,t;y,T)$ of the conditioned process, (i.e. the density of the process $X_t$, conditioned on the event that its density is $f$ at time $T$), is given by

\begin{equation}
    \begin{aligned}
		p^*(x,t;y,T) dy & = \frac{p(x,t;y,T) \, f(y) dy}{p(y,T)} ,\nonumber \\
		                & = \frac{1}{\sqrt{2 \pi (T-t) \sigma^2}} e^{ \textstyle - \frac{(y-x- \mu (T-t))^2}{2(T-t) \sigma^2}} \frac{f(y) dy}{\frac{1}{\sqrt{2 \pi T \sigma^2}}e^{ \textstyle - \frac{(y - \mu T)^2}{2 T \sigma^2}} } \nonumber \\
              & =  \sqrt{\frac{T}{T-t}} e^{ \textstyle - \frac{(y-x- \mu (T-t))^2}{2(T-t) \sigma^2}} e^{\frac{(y - \mu T)^2}{2 T \sigma^2}} f(y)  dy\, .
    \end{aligned}
\end{equation}

\noindent Then,
\begin{equation}
	\pi(x,t) = \int_{-\infty}^{+\infty} p^*(x,t;y,T) \, dy 
              = \sqrt{\frac{T}{T-t}} \int_{-\infty}^{+\infty} e^{ \textstyle - \frac{(y-x- \mu (T-t))^2}{2(T-t) \sigma^2}} e^{\frac{(y - \mu T)^2}{2 T \sigma^2}} f(y) \, dy  \, . \\
\end{equation}
\vspace{0.3cm}
\noindent The logarithmic derivative follows easily: after slightly rearranging the terms, we get
\begin{equation}
\label{drift}
    \begin{aligned}
    \displaystyle \frac{1}{\pi(x,t)} \frac{\partial \pi(x,t)}{\partial x} & = \frac{1}{(T-t)\sigma^2} \frac{ \displaystyle \int_{-\infty}^{+\infty} (y-x- \mu (T-t)) e^{- \textstyle \frac{(y-x- \mu (T-t))^2}{2(T-t) \sigma^2}} e^{ \textstyle \frac{(y - \mu T)^2}{2 T \sigma^2}} f(y)\, dy }{ \displaystyle \int_{-\infty}^{+\infty} e^{ \textstyle - \frac{(y-x- \mu (T-t))^2}{2(T-t) \sigma^2}} e^{ \textstyle \frac{(y - \mu T)^2}{2 T \sigma^2}} f(y) \, dy } \\
	& = - \frac{\mu}{\sigma^2} - \frac{x}{(T-t)\sigma^2} + \frac{1}{(T-t)\sigma^2} 
    \frac{ \displaystyle \int_{-\infty}^{+\infty}  \, y f(y) e^{ \textstyle - \frac{x y}{(T-t) \sigma^2} +\frac{ t y^2}{2 (T-t)T \sigma^2} } \, dy}{ \displaystyle \int_{-\infty}^{+\infty} \, f(y) e^{ \textstyle - \frac{x y}{(T-t) \sigma^2} +\frac{ t y^2}{2 (T-t)T \sigma^2} }\, dy }  \, .
    \end{aligned}
\end{equation}

\noindent Inserting the previous equation into Eq.\eqref{drift_BM_with_drift} leads to
\begin{equation}
	\mu^*(x,t) = \mu +  \frac{\sigma^2}{\pi(x,t)}\frac{\partial \pi(x,t)}{\partial x} = - \frac{x}{(T-t)} + \frac{1}{(T-t)} 
    \frac{ \displaystyle \int_{-\infty}^{+\infty}  \, y f(y) e^{ \textstyle - \frac{x y}{(T-t) \sigma^2} +\frac{ t y^2}{2 (T-t)T \sigma^2} } \, dy}{ \displaystyle \int_{-\infty}^{+\infty} \, f(y) e^{ \textstyle - \frac{x y}{(T-t) \sigma^2} +\frac{ t y^2}{2 (T-t)T \sigma^2} }\, dy } \, .
\end{equation}


\noindent Finally, the SDE for a Brownian motion with constant drift conditioned to be distributed according to a probability distribution function $f$ at time $T$ writes

\begin{equation}
\label{Conditioned_SDE}
	dX^*_t = \frac{1}{(T-t)} \left(  - X^*_t+
    \frac{ \displaystyle \int_{-\infty}^{+\infty} \, y f(y) e^{\textstyle - \frac{y}{(T-t) \sigma^2} \left(\frac{t y}{2 T} -X^*_t \right)} \, dy }{ \displaystyle \int_{-\infty}^{+\infty}  \, f(y) e^{\textstyle -\frac{y}{(T-t) \sigma^2} \left(\frac{t y}{2 T} -X^*_t\right)} \, dy} \right) dt + \sigma dW_t \, .
\end{equation}

\noindent A similar expression, derived using the theory of enlargement of filtration, can be found in~\cite{ref_Baudoin} for a conditioned driftless Brownian motion. The corresponding Langevin equation follows immediately and reads
\begin{equation}
\label{Conditioned_Langevin_SDE}
	\frac{dX^*_t}{dt} = -\frac{X^*_t}{T-t} +\frac{1}{T-t}
    \frac{ \displaystyle \int_{-\infty}^{+\infty} \, y f(y) e^{\textstyle - \frac{y}{(T-t) \sigma^2} \left(\frac{t y}{2 T} -X^*_t \right)} \, dy }{ \displaystyle \int_{-\infty}^{+\infty}  \, f(y) e^{\textstyle -\frac{y}{(T-t) \sigma^2} \left(\frac{t y}{2 T} -X^*_t\right)} \, dy}  + \sigma \eta_t \, .
\end{equation}
\noindent Note that the previous SDE does not depend explicitly on the original $\mu$, although $\mu$ may be implicitly contained in the probability density function $f$. However, when the probability density function $f$ is independent of $\mu$, Eq.\eqref{Conditioned_SDE} shows that the conditioned SDE is also independent of $\mu$. We then have the following result: for a drifted Brownian motion that is conditioned to be distributed  at time $T$ according to a probability function function $f$ independent of the original constant drift, the evolution of the conditioned process is totally independent of the original constant drift, and is given by the stochastic differential equation Eq.\eqref{Conditioned_SDE}. To illustrate this statement, we will now provide a few examples.

\begin{enumerate}
\item By taking $f(x) = \delta(x)$, where $\delta(x)$ is the Dirac delta function, the process is conditioned to end at the origin and as such corresponds to the well known Brownian bridge. From Eq.\eqref{Conditioned_SDE} we immediately get

\begin{equation}
\label{SDE_BB}
	dX^*_t = -\frac{X^*_t}{T-t} dt + \sigma dW_t \, ,
\end{equation}
\noindent which is indeed the stochastic differential equation satisfied by a Brownian bridge~\cite{ref_book_Karlin,ref_Majumdar_Orland,ref_Mazzolo_Jstat}.

\item Similarly, by taking $f(x) = \delta(x-a)$, from Eq.\eqref{Conditioned_SDE} we obtain 
\begin{equation}
\label{SDE_generalized_BB}
	dX^*_t = \frac{a - X^*_t}{T-t} dt + \sigma dW_t \, ,
\end{equation}
\noindent which is the stochastic differential equation satisfied by a Brownian bridge ending at $a$ at time $T$~\cite{ref_book_Karlin,ref_Majumdar_Orland,ref_Mazzolo_Jstat}.

\item In view of the importance of Brownian bridge models for mathematical ecology~\cite{ref_Horne} and  finance~\cite{ref_book_Korn,ref_book_Andersen}, we generalize the previous bridge by authorizing the conditioned process to end at two different locations, say $a$ and $-a$, with possibly two different probabilities, say $\alpha$ and $1-\alpha$. We denote ${\cal{B}}_t$ this process and $\mu_{\cal{B}}(x,t)$ its drift. The density profile at time $T$ is thus $f(x) = \alpha \delta(x-a) + (1-\alpha) \delta(x+a)$: inserting this expression into Eq.\eqref{Conditioned_SDE} we get

\begin{equation}
\label{SDE_two_ends_bridge}
	d{\cal{B}}_t  = -\frac{1}{T-t} \left[a + {\cal{B}}_t -  \frac{2 \, a \, \alpha }{(1 - \alpha) e^{\textstyle -\frac{2 \, a \, {\cal{B}}_t}{(T-t)\sigma^2}} + \alpha}  \right] dt + \sigma dW_t = \mu_{\cal{B}}({\cal{B}}_t ,t) dt + \sigma dW_t \, .
\end{equation}

\noindent The space and time-dependent drift of the process is given by
\begin{equation}
\label{drift_two_ends_bridge}
	\mu_{\cal{B}}(x,t) =  -\frac{1}{T-t} \Bigg( a + x -  \frac{2 \, a \, \alpha }{(1 - \alpha) e^{\textstyle -\frac{2 \, a \, x }{(T-t)\sigma^2}} + \alpha}  \Bigg)  \, ,
\end{equation}

\noindent where $\mu_{\cal{B}}(x,t)$ can be derived from a potential $U_{\cal{B}}(x,t)$, namely $\mu_{\cal{B}}(x,t) = \partial U_{\cal{B}}(x,t)/\partial x$, that is

\begin{equation}
\label{potential_two_ends_bridge}
 	U_{\cal{B}}(x,t) = \frac{x(2 a - x)}{2(T-t)} + \log\left[(1-\alpha)e^{\textstyle-\frac{2 a x}{T-t}} + \alpha\right] \, .
\end{equation}

\noindent This potential is highly dependent of both space and time, and asymmetrical when $\alpha \ne 1/2$. However, as $t$ approaches the final time $T$, $U_{\cal{B}}(x,t)$ converges to a symmetrical shape that is independent of $\alpha$. More precisely, we have
\begin{equation}
\label{limit_potential_two_ends_bridge}
 	\lim_{t \to T} U_{\cal{B}}(x,t) \sim     
     \left\{ 
     	 \begin{aligned} 
		 \frac{x(2 a - x)}{2(T-t)} \mathrm{~if~~} x >0 \\
		 -\frac{x(2 a + x)}{2(T-t)} \mathrm{~if~~} x <0 
     	 \end{aligned}
    	\right. \, ,
\end{equation}

\noindent so that the particle is trapped in a confining quadratic potential either around $a$ or $-a$, with the same intensity. Figure~\ref{fig3} shows the potential at different times: the asymmetry is strongly apparent at the beginning of the process, and then fades away.
\begin{figure}[h]
\centering
\includegraphics[width=5in,height=4.in]{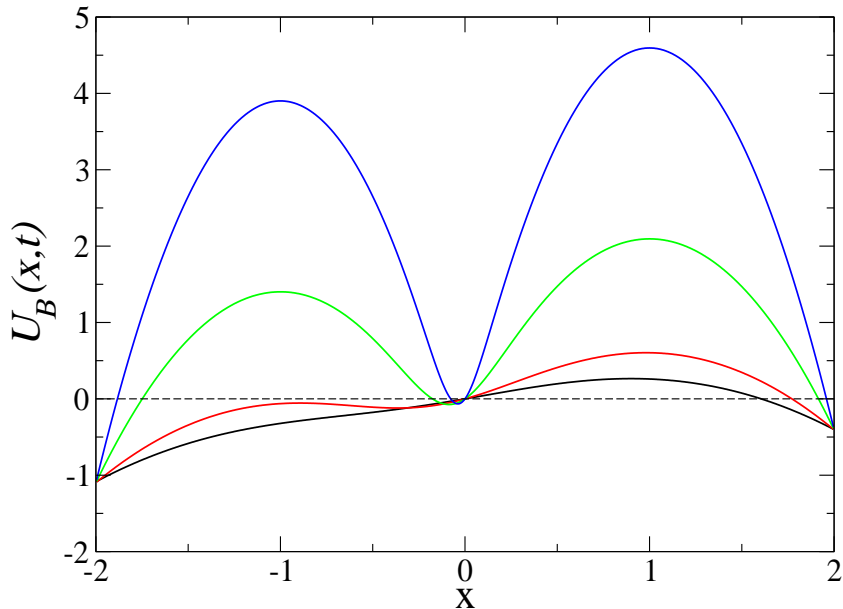}
\setlength{\abovecaptionskip}{15pt} 
\caption{potential $U_{\cal{B}}(x,t)$ as a function of $x$ at various times: $t=0.2$ (black), $t=0.5$ (red), $t=0.8$ (green) and $t=0.9$ (blue). Parameters are $a=1$, $T=1$ and $\alpha = 2/3$. As time increases the asymmetry fades, and the potential becomes symmetrical with respect to $x = 0$ for times near the final time $T=1$.}
\label{fig3}
\end{figure}
In others words, the particle made its choice well before reaching the final time, as one can see in Fig.~\ref{fig4}, where some examples of realizations of the process are shown.
\begin{figure}[h]
\centering
\includegraphics[width=5in,height=4.in]{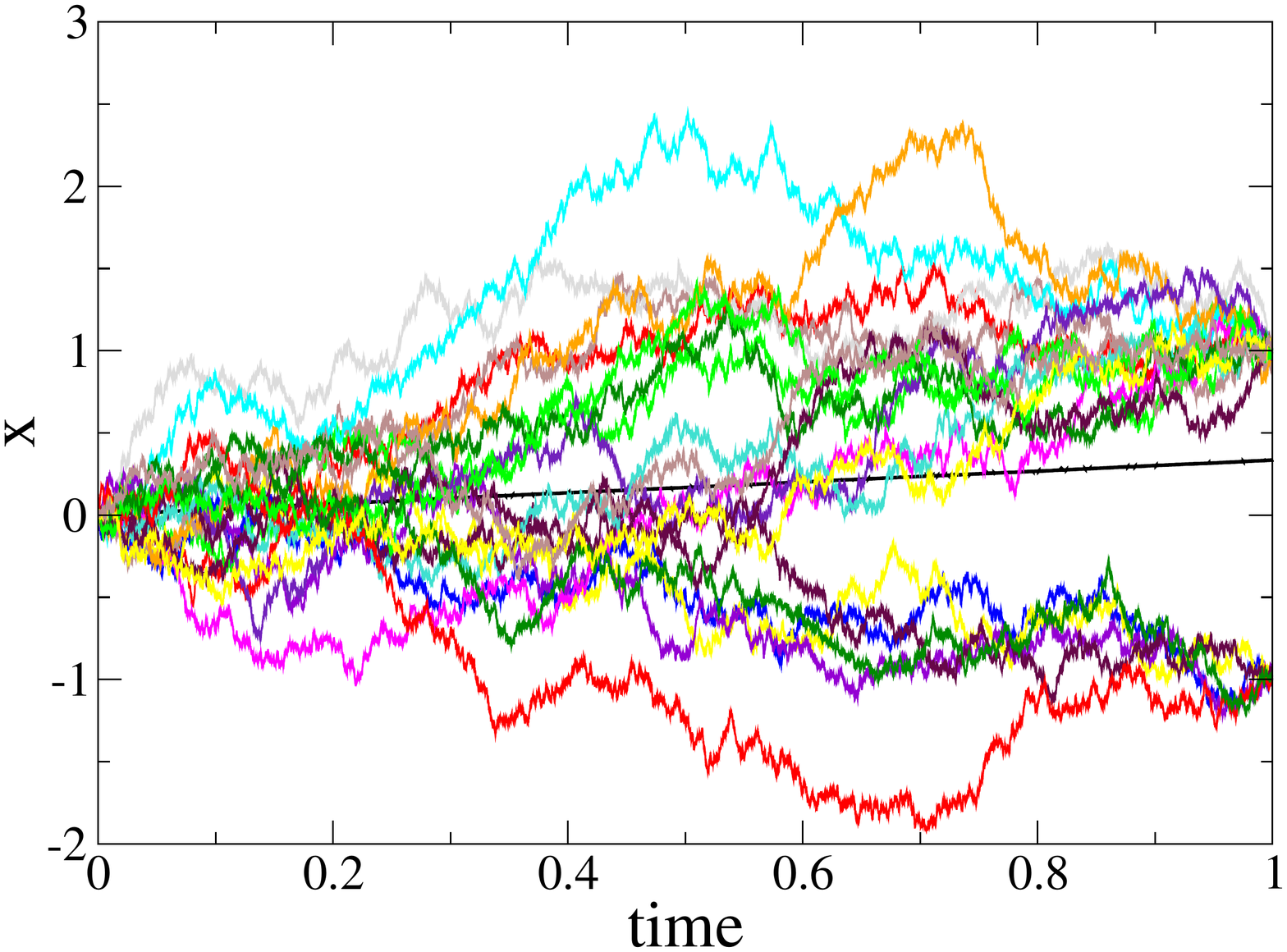}
\setlength{\abovecaptionskip}{15pt} 
\caption{A sample of 20 realizations the two-ends Brownian bridge ending at $a=1$ with probability $\alpha = 2/3$ and at $-a$ with probability $1/3$ at time $T = 1$ (same parameters as for Fig.~\ref{fig3}). The time step used in the discretization is $dt=10^{-4}$. All trajectories generated with different noise histories are statistically independent. The thick black curve is the average profile of the stochastic process obtained by averaging over $10^4$ sample paths. The mean trajectory is a linear function of time given by Eq.\eqref{mean_double_bridge} and corresponds to the mean trajectory of a Brownian bridge ending at $(2\alpha -1) \, a = 1/3$.
}
\label{fig4}
\end{figure}
Indeed, it is  very unlikely that the particle changes its mind approaching the final time. 

\noindent For $\alpha =1/2$ both ends have the same probability of being attained: for this special case, the process will be denoted $\bar{{\cal{B}}_t}$ and its SDE reduces to
\begin{equation}
\label{SDE_two_ends_bridge_symetrical}
	d\bar{{\cal{B}}_t}= \frac{1}{T-t} \left[\bar{{\cal{B}}_t} - a \tanh \left( \frac{a \, \bar{{\cal{B}}_t}}{(T-t)\sigma^2} \right) \right] dt + \sigma dW_t \, .
\end{equation}
\noindent This equation has been previously derived in~\cite{ref_Baudoin}. Generalization to an arbitrary weighted sum of Dirac functions is straightforward, although the involved calculations are cumbersome. 

\item Finally, we consider a Brownian motion with constant drift $\mu$ conditioned to be normally distributed with mean $m T$ and variance $s^2 T$ at time $T$, i.e., $X_T$ has the law $\mathcal{N}(m T,s^2 T)$. Inserting the normal law in Eq.\eqref{Conditioned_SDE}, we obtain 
\begin{equation}
\label{SDE_normal_law}
	dX^*_t = \frac{(s^2 -\sigma^2) X^*_t + m \sigma^2 T}{(s^2 - \sigma^2) t+ \sigma^2 T} dt + \sigma dW_t \, .
\end{equation}
\noindent Naturally, if $m = \mu$ and $s = \sigma$, we recover the original (unconstrained) process,
\begin{equation}
	dX^*_t = \mu \,dt + \sigma dW_t \, = dX_t.
\end{equation}
\noindent Equation~\eqref{SDE_normal_law} is a linear SDE of the form
\begin{equation}
\label{SDE_linear}
	dZ_t = [a_1(t) Z_t + a_2(t)] \,dt + \sigma dW_t \, ,
\end{equation}
\noindent whose solution is a Gaussian process given by~\cite{ref_book_Evans}
\begin{equation}
\label{SDE_linear_solution}
	Z_t = \Phi_{\{t,0\}} \left\{Z_0 + \int_0^t a_2(u)\Phi_{\{u,0\}}^{-1} \, du + \sigma \int_0^t \Phi_{\{u,0\}}^{-1} \, dW_u  \right\} \, ,
\end{equation}
\noindent with $\Phi_{\{t,0\}} = e^{\int_0^t a_1(u) \, du}$. Recalling that $X^*_0 = 0$, from the previous equations we get the solution of Eq.\eqref{SDE_normal_law}, namely,
\begin{equation}
\label{SDE_normal_law_solution}
	X^*_t = m \, t + \sigma \int_0^t  \frac{(s^2 -\sigma^2) t + \sigma^2 T}{(s^2 - \sigma^2) u+ \sigma^2 T} \, dW_u \, .
\end{equation}
\noindent Since the Ito stochastic integral has vanishing expectation, we immediately obtain
\begin{equation}
\label{SDE_normal_law_solution_mean}
	\E[X^*_t] = m \, t  \, .
\end{equation}
\noindent Moreover, for a deterministic (not random) function $A(t)$ Ito's isometry states that~\cite{ref_book_Karlin} 
\begin{equation}
      \E\left[\left(\int_0^t A(u) d W_u \right)^2 \right] = \E\left[\int_0^t A^2(u) du \right] = \int_0^t A^2(u) du \, ,
\end{equation}
\noindent which leads to
\begin{equation}
\label{SDE_normal_law_solution_variance}
	\Var[X^*_t] = \E\left[(X^*_t - m t)^2\right] = \sigma^2 \int_0^t \left(\frac{(s^2 -\sigma^2) t + \sigma^2 T}{(s^2 - \sigma^2) u+ \sigma^2 T} \right)^2  du = \frac{t}{T} [(s^2 -\sigma^2) t + \sigma^2 T] \, .
\end{equation}
\noindent Equations \eqref{SDE_normal_law_solution_mean} and \eqref{SDE_normal_law_solution_variance} fully characterize the conditioned Gaussian diffusion. Note that $\E[X^*_T] = m \, T$ and $\Var[X^*_T] = s^2 T$, as expected.


\end{enumerate}

\section{Two-ends Brownian bridge}
\label{sec4}
\subsection{Standard probabilistic approach}
\label{secStandard}
In the previous section we introduced the two-ends Brownian bridge process ${\cal{B}}_t$ (example iii) as the stochastic process that ends at time $T$ at two different locations: $a$ with probability $\alpha$ and $-a$ with probability $1-\alpha$. This process should not be confused with the sum of two Brownian bridges, which is also a Brownian bridge\footnote{Consider two Brownian bridges of the same length, $B^1_t$ and  $B^2_t$ ending at $a$ and $b$ respectively. The processes satisfy the following SDE, $dB^1_t = \frac{a - B^1_t}{T-t} dt + \sigma dW^1_t $ and $ dB^2_t = \frac{b - B^2_t}{T-t} dt + \sigma dW^2_t $ where $W^1_t$ and $W^2_t$ are two independent Brownian processes. Then $d(B^1_t+B^2_t) =  \frac{a+b - (B^1_t+B^2_t)}{T-t} dt + \sigma d(W^1_t+W^2_t)$. Since $W^1_t+W^2_t$ is a Brownian motion, an immediate consequence is that $B^1_t+B^2_t$ is a Brownian bridge of length $T$ ending at $a+b$.}. For $\alpha = 1/2$, if the two-ends Brownian bridge were a Brownian bridge then it would end at $0$, which is obviously wrong since the two-ends process terminates either at $a$ or $-a$. We will now establish the expression of the probability density function of the two-ends process: its transition density $p_{\cal{B}}(x,t)$ is the unique solution of the Fokker-Planck equation~\cite{ref_book_Schuss}
\begin{equation}
\label{Eq_Fokker_Planck}
	 \frac{\partial p_{\cal{B}}(x,t)}{\partial t} = \frac{\sigma^2}{2}  \frac{\partial^2 p_{\cal{B}}(x,t)}{\partial x^2} - \frac{\partial [\mu_{\cal{B}}(x,t)p_{\cal{B}}(x,t)]}{\partial x} \, ,
\end{equation}
\noindent with the initial delta condition
\begin{equation}
\label{Eq_delta_condition}
	\lim_{t \to 0} p_{\cal{B}}(x,t) = \delta(x).
\end{equation}

\noindent Recall that the drift of two-ends Brownian bridge is given by Eq.\eqref{drift_two_ends_bridge}. Solving the Fokker-Planck equation with such a drift may be quite a challenging task, even in the symmetrical case when $\alpha = 1/2$. However, there exists an alternative strategy for finding the transition probability. First, recall that the transition probability of a Brownian bridge at $a$ at time $T$ is~\cite{ref_Leung}
\begin{equation}
\label{density_Brownian_bridge}
	  p_B(x,t) = \sqrt{\textstyle  \frac{T}{2 \pi t (T-t)\sigma^2} } e^{-\frac{-T \left(\frac{a t}{T} -x \right)^2}{2 t(T-t)\sigma^2} } \, .
\end{equation}
\noindent Next, consider the two-ends process from the conditioned point of view. It is either tied down at $a$ at time $T$ with probability $\alpha$, and in that case its probability density function is that of a Brownian bridge ending at $a$; or it is tied down at $-a$ at time $T$ with probability $1-\alpha$, and in that case its probability density function is that of a Brownian bridge ending at $-a$. Therefore, the probability density function $p_{\cal{B}}(x,t)$ of the two-ends Brownian bridge is the weighted sum of density functions of two Brownian bridges. More precisely,

\begin{equation}
\label{density_double_bridge}
	  p_{\cal{B}}(x,t) =  \alpha \sqrt{\textstyle  \frac{T}{2 \pi t (T-t)\sigma^2} } e^{-\frac{-T \left(\frac{a t}{T} -x \right)^2}{2 t(T-t)\sigma^2} } + (1-\alpha) \sqrt{\textstyle  \frac{T}{2 \pi t (T-t)\sigma^2} } e^{-\frac{-T \left(\frac{a t}{T} +x \right)^2}{2 t(T-t)\sigma^2} } \, .
\end{equation}

\noindent From Eq.\eqref{density_double_bridge} it is straightforward to verify that $p_{\cal{B}}(x,t)$ satisfies the Fokker-Planck equation Eq.\eqref{Eq_Fokker_Planck} with the initial condition $p_{\cal{B}}(x,0) = \delta(x)$. Besides, the mean and variance of the two-ends Brownian bridge process follow easily. Indeed,

\begin{equation}
\label{mean_double_bridge}
	  \E[{\cal{B}}_t] = \int_{-\infty}^{\infty} x \, p_{\cal{B}}(x,t) = \frac{t}{T} (2\alpha -1) \, a  \, ,
\end{equation}

\noindent and

\begin{equation}
\label{variance_double_bridge}
	  \Var[{\cal{B}}_t] = \int_{-\infty}^{\infty} x^2 \, p_{\cal{B}}(x,t) - (\E[{\cal{B}}_t])^2 = \frac{t}{T^2} \left[4 a^2 t (1-\alpha)\alpha  + T(T-t)\sigma^2 \right] \, .
\end{equation}

\noindent Equation~\eqref{mean_double_bridge} shows that to the average of the two-ends Brownian bridge process behaves like that of a Brownian bridge ending at $(2\alpha -1) \, a$ (the weighted sum of two Brownian bridges ending at $a$ and $-a$ with probability $\alpha$ and $1-\alpha$). However, the variance of the process is different from that of a Brownian bridge. For instance, at the final time $T$ we have

\begin{equation}
	  \Var[{\cal{B}}_T] = 4 a^2 \alpha (1 - \alpha) \neq 0 \, ,
\end{equation}

\noindent which is the variance of a Bernoulli process $P(X=a)=\alpha$ and $P(X=-a) = 1-\alpha$, as expected. For a Brownian bridge we would have a zero variance at time $t=T$. Remark also the full coherence between this approach and the SDE of the two-ends Brownian bridge process given in Eq.\eqref{SDE_two_ends_bridge}. Indeed, since $\E[dW_t]=0$, averaging Eq.\eqref{SDE_two_ends_bridge} over the realizations leads to

\begin{equation}
\label{average_SDE_two_ends_bridge}
	d \E[{\cal{B}}_t] = -\frac{1}{T-t} \left\{a + \E[{\cal{B}}_t] -  2 \, a \, \alpha \E\left[\frac{1}{(1 - \alpha) e^{\textstyle -\frac{2 \, a \, {\cal{B}}_t}{(T-t)\sigma^2}} + \alpha}  \right] \right\} dt  .
\end{equation}

\begin{figure}[h]
\centering
\includegraphics[width=5in,height=4.in]{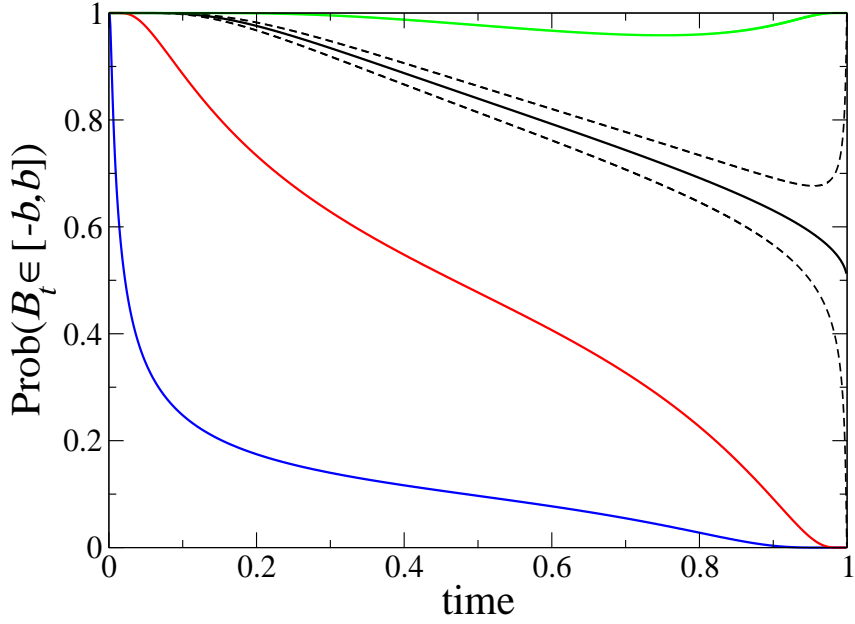}
\setlength{\abovecaptionskip}{15pt} 
\caption{The probability $\mathrm{Prob}\left({\cal{B}}_t \in [-b,b]\right)$ as a function of $t$ for various values of $b$ (other parameters are the same as for the previous plots, namely $a=1$, $T=1$): $b=0.1$ (blue), $b=0.5$ (red), $b=1.5$ (green). The limit case $b=a=1$ is the solid black curve. The dash curve above the limit case corresponds to $b=1.05$ while the dash curve under the limit case is for $b=0.95$. For $b >2$ (not reported on the figure) the probability is always very close to 1.}
\label{fig5}
\end{figure}

\noindent A direct calculation gives
\begin{equation}
	\E\left[\frac{1}{(1 - \alpha) e^{\textstyle -\frac{2 \, a \, {\cal{B}}_t}{(T-t)\sigma^2}} + \alpha}  \right] = \int_{-\infty}^{\infty} \frac{p_{\cal{B}}(x,t)}{(1 - \alpha) e^{\textstyle -\frac{2 \, a \, {\cal{B}}_t}{(T-t)\sigma^2}} + \alpha}  \, dx   = 1 \, ,
\end{equation}
\noindent and Eq.\eqref{average_SDE_two_ends_bridge} reduces to a simple linear first-order differential equation, i.e.,

\begin{equation}
	\frac{d \E[{\cal{B}}_t]}{dt} = -\frac{1}{T-t} \left[ \E[{\cal{B}}_t] + a (1- 2 \, \alpha) \right]   \, ,
\end{equation}

\noindent whose solution with the initial condition $\E[{\cal{B}}_0] =0 $ is precisely  Eq.\eqref{mean_double_bridge}. In order to better understand the behavior of the process, we can look at the probability of finding the particle in an interval $[-b; b]$ at different times. This quantity is easily derived from the previous density Eq.\eqref{density_double_bridge}, indeed

\begin{equation}
\begin{aligned}
	 \mathrm{Prob}\left({\cal{B}}_t \in [-b,b]\right) & = \int_{-b}^{b} p_{\cal{B}}(x,t) \, dx \\ 
                                                       & = \frac{1}{2} \left[
 \mathrm{erf} \left( \frac{at + bT}{\sqrt{2 \,t \,T (T-t) \sigma^2}} \right)  - \mathrm{erf} \left( \frac{at - bT}{\sqrt{2 \,t \,T (T-t) \sigma^2}} \right) \right] .
\end{aligned}
\end{equation}
\noindent It is worth noting that the probability is independent of $ \alpha$. Besides, $\mathrm{Prob}\left({\cal{B}}_0 \in [-b,b]\right) = 1$ (the process starts in the interval) and $\mathrm{Prob}\left({\cal{B}}_T \in [-b,b]\right) = 1 $ if $b >a$ and $0$ if $b<a$. In the limit case where $b = a$, the process has fifty percent probability of being in the interval at the final time $T$, as shown in Fig.\ref{fig5}.

We further characterize this process by computing the conditional probability $P({\cal{B}}_T=a | {\cal{B}}_t=x)$ of hitting $a$ at time $T$,  knowing that the process was in $x$ at a time $t<T$. In the absence of constraint, the probability of hitting a small interval $da$ around $a$ at the final time $T$ for a Brownian motion with constant drift $\mu$ is
\begin{equation}
	\frac{1}{\sqrt{2 \pi T \sigma^2}} e^{ \textstyle - \frac{\left(a-\mu T\right)^2}{2 T \sigma^2} } da \, .  
\end{equation}

\noindent Similarly, the probability of hitting a small interval $da$ around $-a$ at the final time $T$ is $ \frac{1}{\sqrt{2 \pi T \sigma^2}} e^{ \textstyle - \frac{\left(-a-\mu T\right)^2}{2 T \sigma^2} } da $. For the constrained process, we only consider trajectories that reach these two intervals with probabilities $\alpha$ and $1-\alpha$. Under this constraint, the probability of hitting a small interval $da$ around $a$ at the final time $T$ is weighted by a factor $\alpha/(\frac{1}{\sqrt{2 \pi T \sigma^2}} e^{ \textstyle - \frac{\left(a-\mu T\right)^2}{2 T \sigma^2} } da)   
$, and similarly by a factor $(1-\alpha)/(\frac{1}{\sqrt{2 \pi T \sigma^2}} e^{ \textstyle - \frac{\left(-a-\mu T\right)^2}{2 T \sigma^2} } da )$ for the small interval $da$ around $-a$. Now, consider a trajectory starting at $x$ at time $t$. For such a process, up to a normalization constant, the probabilities of reaching $\pm a$ are respectively given by

\begin{equation}
  \left\{
       \begin{aligned}
       & \mathrm{~~~Prob}\left( {\cal{B}}_T \in [a,a+da] | {\cal{B}}_t =x \right)  & \propto && \frac{e^{ \textstyle - \frac{\left(x-\mu (T-t) -a\right)^2}{2(T-t) \sigma^2} } \, da }{\sqrt{2 \pi (T-t) \sigma^2}}  
\times  \frac{\alpha}{\frac{1}{\sqrt{2 \pi T \sigma^2}} e^{ \textstyle - \frac{\left(a-\mu T\right)^2}{2 T \sigma^2} } da}   \\
       & \mathrm{Prob}\left( {\cal{B}}_T \in [-a,-a+da] | {\cal{B}}_t =x \right) & \propto && \frac{e^{\textstyle - \frac{\left(x-\mu (T-t) +a\right)^2}{2(T-t) \sigma^2} }\, da}{\sqrt{2 \pi (T-t) \sigma^2}}  
 \times  \frac{1-\alpha}{\frac{1}{\sqrt{2 \pi T \sigma^2}} e^{ \textstyle - \frac{\left(-a-\mu T\right)^2}{2 T \sigma^2} } da} \, . \nonumber
       \end{aligned}
   \right.
\end{equation}

\noindent After normalizing the probabilities, we get

\begin{equation}  
\label{proba_Pa}
   \left\{
       \begin{aligned}
	      P({\cal{B}}_T=a | {\cal{B}}_t=x) & = \frac{\alpha}{\alpha + \displaystyle (1-\alpha) \, e^{ \textstyle - \frac{2 a x}{(T-t) \sigma^2}} } \\   
		 P({\cal{B}}_T=-a | {\cal{B}}_t=x) & = \frac{1-\alpha}{(1-\alpha) + \alpha \, \displaystyle e^{ \textstyle \frac{2 a x}{(T-t) \sigma^2}} }\, .    
       \end{aligned} 
   \right.
\end{equation}

\noindent Remark that, since the two-ends Brownian bridge process is independent of $\mu$, these probabilities do not depend of the original drift $\mu$, as expected. If $\alpha = 1/2$ (symmetrical case) then

\begin{equation}
\label{Proba_a}
   \left\{
       \begin{aligned}
	P({\cal{B}}_T=a | {\cal{B}}_t=x)  &= \frac{1}{1+ e^{ \textstyle - \frac{2 a x}{(T-t) \sigma^2}} } &&= \frac{1}{2} \left(1 + \tanh \left( \frac{a x}{(T-t) \sigma^2} \right) \right) \\
	P({\cal{B}}_T=-a | {\cal{B}}_t=x) &= \frac{1}{1+ e^{ \textstyle  \frac{2 a x}{(T-t) \sigma^2}} } &&= \frac{1}{2} \left(1 - \tanh \left( \frac{a x}{(T-t) \sigma^2} \right) \right) 
  	\, .    
       \end{aligned} 
   \right.
\end{equation}

\noindent Our expressions correct by a factor 2 the exponential as given in~\cite{ref_Baudoin} for the symmetrical driftless process. Monte Carlo simulations confirm our findings. Of course, when $x=0$, Eqs.\eqref{Proba_a} give $P({\cal{B}}_T=a | {\cal{B}}_t=0) = \alpha$ and $P({\cal{B}}_T=-a | {\cal{B}}_t=0) = 1-\alpha$, as expected. Remark also that when $t \to T$ the conditioned probabilities depend only on the sign of $x$. More precisely, $\lim_{t \to T} P({\cal{B}}_T=a | {\cal{B}}_t=x) = 1 $ if $x>0$, and $0$ otherwise, meaning that the process strongly feels the boundaries as the current time approaches the final time. On the contrary, at the beginning of the process, the drift term behaves as

\begin{equation}
\label{drift_two_ends_bridge_approx}
	\lim_{x,t \to 0} \mu_{\cal{B}}(x,t) \sim  \frac{2 \alpha -1}{T}  +o(x) \, ,
\end{equation}

\noindent so that the process feels only slightly the asymmetry, through a positive drift if $\alpha >1/2$ and a negative drift if $\alpha <1/2$, which is again coherent with the results shown in Fig.~\ref{fig3} (black curve).\\

Similarly, based on the distribution of the first-passage time of the Brownian bridge we can obtain the distribution of the first-passage time of the two-ends Brownian bridge. To this aim, consider a positive level $\beta$ and let $T_{\beta} = \inf\{s<t: B_s = \beta\}$ be the first time the Brownian bridge reaches this level. For such a process, recall that for $\beta >0$ we have~\cite{ref_Beghin} 

\begin{equation}
\label{Proba_T_beta_BB}
  \begin{aligned}
	& P( T_{\beta} \le t  \,| B_T=a)  = P\left(\max_{0 \le s \le t} B_s \ge \beta \,| B_T=a \right) \\
     						    & = 
       \left\{        
       \begin{aligned}
                 & e^{ \textstyle -\frac{2 \beta (\beta -a)}{T \sigma^2} }\int_{-\infty}^{\frac{2 \beta t - a t - \beta T}{\sigma \sqrt{t T (T-t)}}}  \frac{e^{ \textstyle - \frac{y^2}{2}}}{\sqrt{2 \pi}} \, dy  + \int_{\frac{\beta T - a t}{\sigma \sqrt{t T (T-t)}}}^{\infty} \frac{e^{ \textstyle - \frac{y^2}{2}}}{\sqrt{2 \pi}} \, dy   \qquad \mathrm{~for~~} 0<t<T  \\
                 &
                 \left\{        
                 \begin{aligned}
				 & e^{ \textstyle -\frac{2 \beta (\beta -a)}{T \sigma^2} }+  2  \left( 1- e^{ \textstyle -\frac{2 \beta (\beta -a)}{T \sigma^2} } \right) \int_{\frac{\beta - a}{ \sigma \sqrt{t - T}}}^{\infty} \frac{e^{ \textstyle - \frac{y^2}{2}}}{\sqrt{2 \pi}} \, dy && \qquad \mathrm{for~~} t>T \mathrm{~~and ~~} \beta > a \\
                     & 1 && \qquad \mathrm{for~~} t>T \mathrm{~~and ~~} \beta < a \, .
                \end{aligned} 
                \right.
       \end{aligned} 
       \right.
  \end{aligned}
\end{equation}

\noindent As in the previous section, we are interested in events occurring during the time interval $[0,T]$. During this interval, for a given $\beta > a$, Eq.\eqref{Proba_T_beta_BB}  indicates that there is a non-zero probability that the level $\beta$ is not hit and the (conditional) first-exit time density probability should be properly normalized\footnote{The normalization constant is obtained by calculating the limit: $\lim_{t \to T} P( T_{\beta} \le t  \,| B_T=a) $ and is equal to $ e^{ \textstyle - \frac{2 \beta (\beta -a)}{T \sigma^2} } $ when $\beta > a$ and $1$ otherwise.}. The density function $g_{\beta}(t)$ of the first hitting time, conditioned to the particle actually reaching the level $\beta$ during $[0,T]$, is given by

\begin{equation}
\label{density_T_beta_BB}
	g_{\beta}(t) = \frac{\partial}{\partial t} P( T_{\beta} \le t | B_T=a)  =
                 \left\{        
                 \begin{aligned}
                        & \textstyle{\beta \sqrt{\frac{T}{2 \pi t^3 (T-t)\sigma^2}} } e^{ \textstyle - \frac{(\beta T - a t)^2 }{2 t T (T-t)\sigma^2} } e^{ \textstyle \frac{2 \beta (\beta -a)}{T \sigma^2} }  && \mathrm{for~~}  \beta > a\\ 
                        & \textstyle{\beta \sqrt{\frac{T}{2 \pi t^3 (T-t)\sigma^2}} } e^{ \textstyle - \frac{(\beta T - a t)^2 }{2 t T (T-t)\sigma^2} }   && \mathrm{for~~} \beta < a \, .
                \end{aligned} 
                \right. 
\end{equation}

\noindent With these results at hand, it is a simple matter to obtain the probability density function ${\cal{G}}_{\beta}(t)$ of the first hitting time for the two-ends Brownian bridge, namely

\begin{equation}
\label{density_T_beta_2_ends_BB}
	{\cal{G}}_{\beta}(t)  =
                 \left\{        
                 \begin{aligned}
                        & \textstyle{\beta \sqrt{\frac{T}{2 \pi t^3 (T-t)\sigma^2}} } \left[ \frac{ \alpha e^{ \textstyle - \frac{(\beta T - a t)^2 }{2 t T (T-t)\sigma^2} }+  (1-\alpha)  e^{ \textstyle - \frac{(\beta T + a t)^2 }{2 t T (T-t)\sigma^2} } }
{\alpha e^{ \textstyle - \frac{2 \beta (\beta -a)}{T \sigma^2} } +(1-\alpha) e^{ \textstyle - \frac{2 \beta (\beta +a)}{T \sigma^2} }  } \right] && \mathrm{for~~} \beta > a\\
                        & \textstyle{\beta \sqrt{\frac{T}{2 \pi t^3 (T-t)\sigma^2}} } \left[ \frac{ \alpha e^{ \textstyle - \frac{(\beta T - a t)^2 }{2 t T (T-t)\sigma^2} }+  (1-\alpha)  e^{ \textstyle - \frac{(\beta T + a t)^2 }{2 t T (T-t)\sigma^2} } }
{\alpha  +(1-\alpha) e^{ \textstyle - \frac{2 \beta (\beta +a)}{T \sigma^2} }  } \right] && \mathrm{for~~} \beta < a \, .
                \end{aligned} 
                \right. 
\end{equation}

\begin{figure}[h]
\centering
\includegraphics[width=5in,height=4.in]{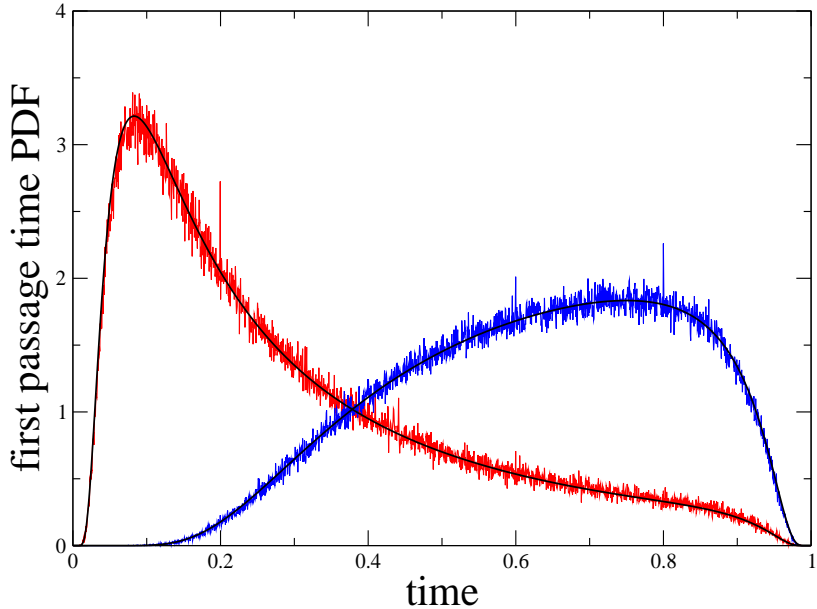}
\setlength{\abovecaptionskip}{15pt} 
\caption{Probability density function of the first hitting time for the two-ends Brownian bridge: in red the  simulations corresponding to $\beta = a/2$ and in blue those for $\beta = 3a/2$. Black curves correspond to the exact expressions given by Eq.\eqref{density_T_beta_2_ends_BB} .}
\label{fig6}
\end{figure}

\noindent The behavior of the probability density function of the first hitting time varies considerably depending on whether $\beta$ is higher or lower than $a$, as shown in Fig.~\ref{fig6}. 

\subsection{Martingale approach}
\label{secMartingale}
In this section we explore the two-ends Brownian bridge process by resorting to martingale techniques. 
The pioneering works of Mark Kac have established the deep connection between pure probabilistic quantities and (partial) differential equations, via the celebrated Feynman-Kac formula~\cite{ref_Kac_original,ref_book_Kac}. Since then, the Feynman-Kac formalism has been successfully applied to various kind of Brownian functionals (see for instance~\cite{ref_Majumdar_CS} for a recent review) as well as for other Markovian continuous-time processes~\cite{ref_ZDM_PRE_2011a} and non-Markovian processes~\cite{ref_Barkai,ref_Turgeman}. Martingales are fundamental objects for the analysis of stochastic processes, which also considerably simplify the calculations of probabilistic quantities such as expectation, conditional probability or first exit times~\cite{ref_book_Williams}.
For instance, the classical result concerning the probability that a Brownian motion with positive constant drift eventually hits the origin  (which is needed for the derivation of the first-passage duality~\cite{ref_Krapivsky}) can be obtained by standard probabilistic methods, as done in~\cite{ref_Krapivsky,ref_book_Redner} or by a very elegant method (almost without calculations) by introducing an appropriate martingale as described in Lawler's book~\cite{ref_book_Lawler}. In this paragraph, based on the martingale technique, we will derive the partial differential equation satisfied by the conditioned probability $P({\cal{B}}_T=a | {\cal{B}}_t=x)$ of hitting $a$ at time $T$ knowing that the two-ends Brownian bridge process (introduced in the previous section) was in $x$ at time $t<T$. For this purpose, consider the final event: $\{{\cal{B}}_T = a \}$. The probability $P({\cal{B}}_T=a | {\cal{B}}_t=x)$, denoted $\phi(x,t)$, can now be expressed as a conditional expectation, namely,

\begin{equation}
		\phi(x,t) = P({\cal{B}}_T=a | {\cal{B}}_t=x) = \E\left[\mathbbm{1}_{\{{\cal{B}}_T = a \}} | {\cal{B}}_t=x \right] \, ,
\end{equation}
\noindent where $\mathbbm{1}_{\{{\cal{B}}_T = a \}} $ is the indicator function of the event $\{{\cal{B}}_T = a \}$. Closely following~\cite{ref_book_Lawler}, we introduce

\begin{equation}
		M_t = \E\left[\mathbbm{1}_{\{{\cal{B}}_T = a \}} | \mathcal{F}_t \right] \, ,
\end{equation}
\noindent where $\mathcal{F}_t$ is the filtration (i.e., the information at time $t$) generated by the process ${\cal{B}}_t$. Then, using the tower property of conditional expectations $(s < t)$~\cite{ref_book_Wiersema}, we get

\begin{equation}
	    \E\left[M_t | \mathcal{F}_s \right] =   \E\left[ \E\left[\mathbbm{1}_{\{{\cal{B}}_T = a \}} | \mathcal{F}_t \right] | \mathcal{F}_s \right] = \E\left[  \mathbbm{1}_{\{{\cal{B}}_T = a \}}    | \mathcal{F}_s \right]  = M_s     \, .
\end{equation}

\noindent This relationship shows that $M_t$ is a martingale. Besides, due to the Markov property of the diffusion process ${\cal{B}}_t$, the amount of information at time $t$ is just the location of the process (i.e., ${\cal{B}}_t=x$): $\phi({\cal{B}}_t,t)$ is therefore a martingale (note that, since $\E\left[\mathbbm{1}_{\{{\cal{B}}_T = a \}} | {\cal{B}}_t=x \right] \in [0,1]$, it is also a bounded martingale). Now, applying Ito's formula to $\phi({\cal{B}}_t,t)$ gives~\cite{ref_book_Karlin}

\begin{equation}
    \label{eq_SDE_dphi}
       d\phi({\cal{B}}_t,t)  = \frac{\partial \phi({\cal{B}}_t,t)}{\partial t} dt + \frac{\partial \phi({\cal{B}}_t,t)}{\partial x} d{\cal{B}}_t + \frac{\sigma^2}{2} \frac{\partial^2 \phi({\cal{B}}_t,t)}{\partial x^2} dt \, .   
\end{equation}

\noindent Inserting Eq.\eqref{SDE_two_ends_bridge} into the equation~\eqref{eq_SDE_dphi} yields

\begin{equation}
    d\phi({\cal{B}}_t,t)  = \left[\frac{\partial \phi({\cal{B}}_t,t)}{\partial t} + \mu_{\cal{B}}(x,t) \frac{\partial \phi({\cal{B}}_t,t)}{\partial x} + \frac{\sigma^2}{2} \frac{\partial^2 \phi({\cal{B}}_t,t)}{\partial x^2} \right] dt  + \sigma \frac{\partial \phi({\cal{B}}_t,t)}{\partial x} dW_t \, .
\end{equation}



\noindent Since a martingale is a stochastic process which has a zero drift at all times~\cite{ref_book_Lawler}, the term $\propto dt$ in the previous equation must vanish, and the function $\phi(x,t)$ must satisfy the partial differential equation

\begin{equation}
    \label{eq_phi_generale}
    - \frac{\partial \phi(x,t)}{\partial t} = \mu_{\cal{B}}(x,t) \frac{\partial \phi(x,t)}{\partial x} + \frac{\sigma^2}{2} \frac{\partial^2 \phi(x,t)}{\partial x^2}  \, ,
\end{equation}

\noindent which yields
\begin{equation}
    \label{eq_phi}
    \frac{\partial \phi(x,t)}{\partial t} = \frac{1}{T-t} \bigg(a + x -  \frac{2 \, a \, \alpha }{(1 - \alpha) e^{\textstyle -\frac{2 \, a \, x}{(T-t)\sigma^2}} + \alpha}  \bigg) \frac{\partial \phi(x,t)}{\partial x} - \frac{\sigma^2}{2} \frac{\partial^2 \phi(x,t)}{\partial x^2}  \, ,
\end{equation}

\noindent with the boundary condition $\phi(0,0) = P({\cal{B}}_T=a | {\cal{B}}_0=0) = \alpha$. The previous equation can be solved exactly with the boundary condition and the additional constraint that $\phi(x,t) \ge 0$. This would nevertheless require the same amount of work as for solving the Fokker-Planck equation Eq.\eqref{Eq_Fokker_Planck}. It is straightforward to verify that Eq.\eqref{proba_Pa} is the unique solution of Eq.\eqref{eq_phi} with the appropriate boundary conditions. Remark that, up to the minus sign, the partial differential equation for the conditioned probability looks like a Kolmogorov backward differential equation. \\
\noindent In the symmetrical case, when $\alpha = 1/2$, the equation~\eqref{eq_phi} becomes

\begin{equation}
    \label{eq_phi_sym}
    \frac{\partial \phi(x,t)}{\partial t} = \frac{1}{T-t} \bigg(a \tanh \left( \frac{a \, x}{(T-t)\sigma^2} \right) -x \bigg)
    \frac{\partial \phi(x,t)}{\partial x} - \frac{\sigma^2}{2} \frac{\partial^2 \phi(x,t)}{\partial x^2}  \, .
\end{equation}

\noindent From the expression of $\mu_{\cal{B}}(x,t)$ (see Eq.\eqref{drift_two_ends_bridge}) we also remark that the space-time dependency appears through the variable $x/(T-t)$, and we can guess that the solution $\phi(x,t)$ of Eq.\eqref{eq_phi} shares the same dependency. Seeking a solution of the form $\phi(x,t) = \varphi(x/(T-t))$, we have $\partial \varphi/ \partial t = x/(T-t) \partial \varphi/ \partial x$, and Eq.\eqref{eq_phi} reduces to an ordinary second-order differential equation, namely,

\begin{equation}
    \label{eq_varphi}
    \frac{1}{T-t} \bigg(a -  \frac{2 \, a \, \alpha }{(1 - \alpha) e^{\textstyle -\frac{2 \, a \, x}{(T-t)\sigma^2}} + \alpha}  \bigg)\varphi' - \frac{\sigma^2}{2} \varphi'' = 0 \, .
\end{equation}

\noindent Equation~\eqref{eq_varphi} can be easily integrated with the initial condition $\varphi(0) = \alpha$ and $\lim_{x \to \infty}\varphi(x) = 1$ (or equivalently $\lim_{x \to -\infty}\varphi(x)  = 0$, meaning that if the particle starts infinitely far from the two targets it will hit the closest one with certainty) and we recover the expression given by Eq.\eqref{proba_Pa}.\\

As a final remark, observe that the approach developed in this paragraph is general, and Eq.\eqref{eq_phi_generale} is valid for any drift (here we have considered the particular form $\mu_{\cal{B}}(x,t)$). Therefore, the equation

\begin{equation}
   - \frac{\partial \phi(x,t)}{\partial t} = \mu(x,t) \frac{\partial \phi(x,t)}{\partial x} + \frac{\sigma^2}{2} \frac{\partial^2 \phi(x,t)}{\partial x^2} 
\end{equation}

\noindent along with its boundary conditions can be seen as a 
tool to compute conditional probabilities on the final state of a diffusion process driven by the SDE, $dX_t = \mu(x,t) dt+ \sigma dW_t $.

\section{Conclusions}
\label{sec_Conclusion}
Conditioned stochastic processes have often striking priorities. In this paper, we first sought to understand the recent phenomenon of first-passage duality through the effective Langevin equation approach. This led us to answer an important question closely related to the first-passage duality: what kind of constraints on a Brownian motion with constant drift leads to a stochastic process that is independent of the initial drift? After observing that a sufficient condition is to have a final density distribution independent of the initial drift, we have investigated the properties of a generalized Brownian bridge that can end at two different locations (with probabilities that may also be different). We believe that this process and its generalizations will play an important role in mathematical ecology and finance, notably by replacing a succession of Brownian bridges by the two (or more)-ends process introduced in this paper.



\newpage

\end{document}

%% file: brownianhitting.tex
\begingroup
  \makeatletter
  \providecommand\color[2][]{%
    \GenericError{(gnuplot) \space\space\space\@spaces}{%
      Package color not loaded in conjunction with
      terminal option `colourtext'%
    }{See the gnuplot documentation for explanation.%
    }{Either use 'blacktext' in gnuplot or load the package
      color.sty in LaTeX.}%
    \renewcommand\color[2][]{}%
  }%
  \providecommand\includegraphics[2][]{%
    \GenericError{(gnuplot) \space\space\space\@spaces}{%
      Package graphicx or graphics not loaded%
    }{See the gnuplot documentation for explanation.%
    }{The gnuplot epslatex terminal needs graphicx.sty or graphics.sty.}%
    \renewcommand\includegraphics[2][]{}%
  }%
  \providecommand\rotatebox[2]{#2}%
  \@ifundefined{ifGPcolor}{%
    \newif\ifGPcolor
    \GPcolorfalse
  }{}%
  \@ifundefined{ifGPblacktext}{%
    \newif\ifGPblacktext
    \GPblacktexttrue
  }{}%
  \let\gplgaddtomacro\g@addto@macro
  \gdef\gplbacktext{}%
  \gdef\gplfronttext{}%
  \makeatother
  \ifGPblacktext
    \def\colorrgb#1{}%
    \def\colorgray#1{}%
  \else
    \ifGPcolor
      \def\colorrgb#1{\color[rgb]{#1}}%
      \def\colorgray#1{\color[gray]{#1}}%
      \expandafter\def\csname LTw\endcsname{\color{white}}%
      \expandafter\def\csname LTb\endcsname{\color{black}}%
      \expandafter\def\csname LTa\endcsname{\color{black}}%
      \expandafter\def\csname LT0\endcsname{\color[rgb]{1,0,0}}%
      \expandafter\def\csname LT1\endcsname{\color[rgb]{0,1,0}}%
      \expandafter\def\csname LT2\endcsname{\color[rgb]{0,0,1}}%
      \expandafter\def\csname LT3\endcsname{\color[rgb]{1,0,1}}%
      \expandafter\def\csname LT4\endcsname{\color[rgb]{0,1,1}}%
      \expandafter\def\csname LT5\endcsname{\color[rgb]{1,1,0}}%
      \expandafter\def\csname LT6\endcsname{\color[rgb]{0,0,0}}%
      \expandafter\def\csname LT7\endcsname{\color[rgb]{1,0.3,0}}%
      \expandafter\def\csname LT8\endcsname{\color[rgb]{0.5,0.5,0.5}}%
    \else
      \def\colorrgb#1{\color{black}}%
      \def\colorgray#1{\color[gray]{#1}}%
      \expandafter\def\csname LTw\endcsname{\color{white}}%
      \expandafter\def\csname LTb\endcsname{\color{black}}%
      \expandafter\def\csname LTa\endcsname{\color{black}}%
      \expandafter\def\csname LT0\endcsname{\color{black}}%
      \expandafter\def\csname LT1\endcsname{\color{black}}%
      \expandafter\def\csname LT2\endcsname{\color{black}}%
      \expandafter\def\csname LT3\endcsname{\color{black}}%
      \expandafter\def\csname LT4\endcsname{\color{black}}%
      \expandafter\def\csname LT5\endcsname{\color{black}}%
      \expandafter\def\csname LT6\endcsname{\color{black}}%
      \expandafter\def\csname LT7\endcsname{\color{black}}%
      \expandafter\def\csname LT8\endcsname{\color{black}}%
    \fi
  \fi
  \setlength{\unitlength}{0.0500bp}%
  \begin{picture}(7200.00,5040.00)%
    \gplgaddtomacro\gplbacktext{%
      \csname LTb\endcsname%
      \put(726,1679){\makebox(0,0)[r]{\strut{}-0.5}}%
      \put(726,3227){\makebox(0,0)[r]{\strut{} 0}}%
      \put(726,4775){\makebox(0,0)[r]{\strut{} 0.5}}%
      \put(858,220){\makebox(0,0){\strut{} 0}}%
      \put(2047,220){\makebox(0,0){\strut{} 0.2}}%
      \put(3236,220){\makebox(0,0){\strut{} 0.4}}%
      \put(4425,220){\makebox(0,0){\strut{} 0.6}}%
      \put(5614,220){\makebox(0,0){\strut{} 0.8}}%
      \put(6803,220){\makebox(0,0){\strut{} 1}}%
      \put(2523,130){\makebox(0,0)[l]{\strut{}\Huge $t$ \normalsize}}%
      \put(472,1059){\makebox(0,0)[l]{\strut{}\Huge $x$ \normalsize}}%
      \put(6660,-177){\makebox(0,0)[l]{\strut{}\Huge $T_a$ \normalsize}}%
      \put(6922,4837){\makebox(0,0)[l]{\strut{}\Huge $a$ \normalsize}}%
      \put(6922,4496){\makebox(0,0)[l]{\strut{}\Huge $a$-$\varepsilon$ \normalsize}}%
    }%
    \gplgaddtomacro\gplfronttext{%
    }%
    \gplbacktext
    \put(0,0){\includegraphics{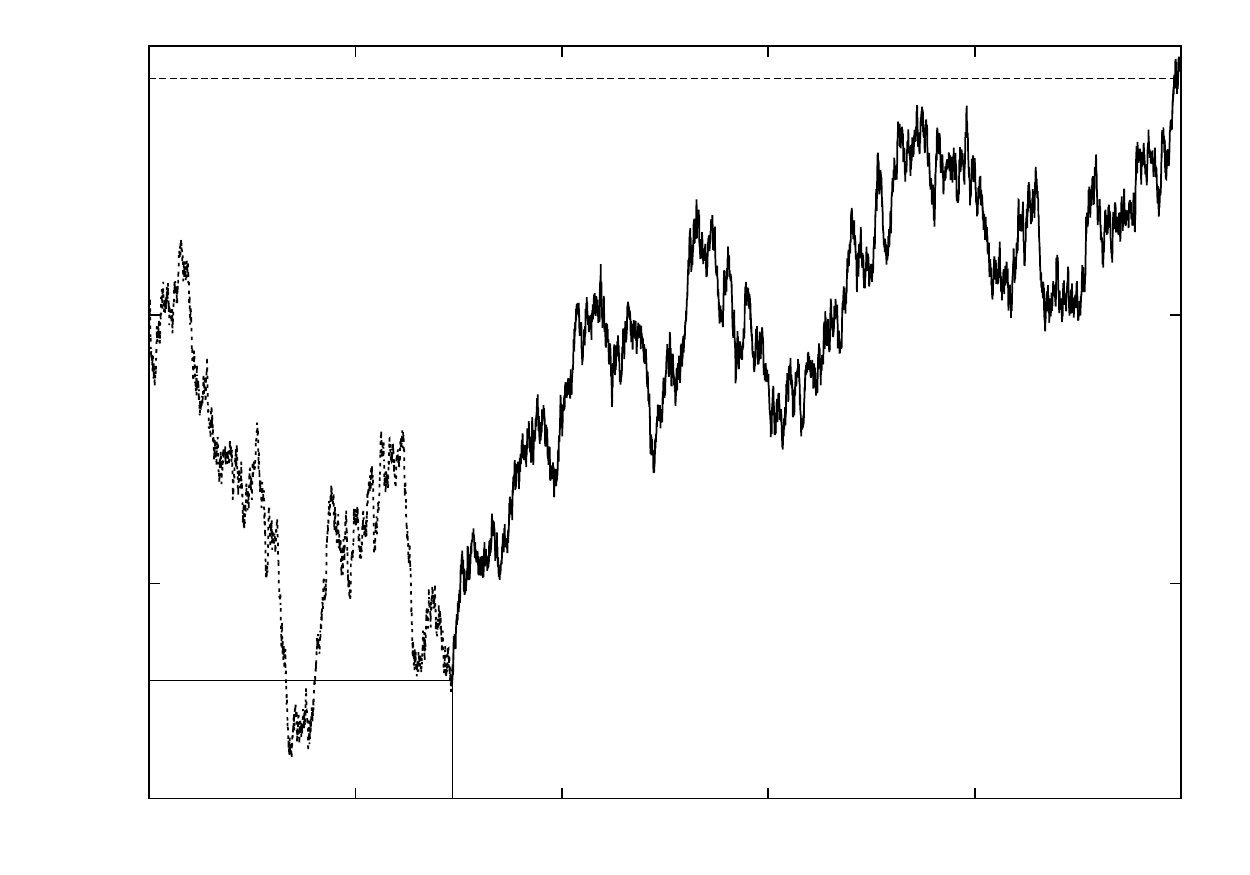}}%
    \gplfronttext
  \end{picture}%
\endgroup